\documentclass[symmetry,article,accept,moreauthors,pdftex]{Definitions/mdpi} 
\firstpage{1} 
\pubvolume{00}
\issuenum{1}
\articlenumber{5}
\pubyear{2020}
\copyrightyear{2020}
\history{Received: 25 August 2020; Accepted: 15 October 2020; Published: date}
\updates{yes} 

\usepackage{color}

 \usepackage{xcolor, colortbl}
\usepackage[normalem]{ulem}
\usepackage{longtable}
\usepackage{threeparttable}
\usepackage{booktabs,multirow}
\usepackage{attrib}
\usepackage{epstopdf}
\usepackage{subfigure}
\usepackage{mathcomp}
\usepackage{amsfonts}
\usepackage{amssymb}
 \usepackage{amsmath}

\Title{An Anisotropic Model for the Universe} 

\renewcommand{\hl}{}
\Author{Morgan Le~Delliou 
 $^{1,2}$, Maksym Deliyergiyev $^{3}$, and \hl{Antonino Del Popolo} 
 $^{4,5,6}$*}

\AuthorNames{Morgan Le~Delliou, Maksym Deliyergiyev and Antonino Del Popolo}

\address{%
$^{1}$ \quad Institute of Theoretical Physics, School of Physical Science and
Technology, Lanzhou University, No.222, South Tianshui Road, Lanzhou 730000,
Gansu, China; delliou@lzu.edu.cn or Morgan.LeDelliou.IFT@gmail.com\\ 
$^{2}$ \quad Instituto de Astrofsica e Cincias do Espao, Universidade de Lisboa, Faculdade de Cincias, Ed. C8, Campo Grande, 1769-016 Lisboa, Portugal\\
$^{3}$ \quad D\'epartement de Physique Nucl\'eaire et Corpusculaire, University of
Geneva, CH-1211 Geneve, Switzerland; maksym.deliyergiyev@unige.ch\\
$^{4}$ \quad Dipartimento di Fisica e Astronomia, University Of Catania, Viale
Andrea Doria, 95125 Catania, Italy\\
$^{5}$ \quad Institute of Astronomy, Russian Academy of Sciences, , Pyatnitskaya
str., 48, {119017 Moscow,} 
 Russia\\
$^{6}$ \quad INFN sezione di Catania, Via S. Sofia 64, I-95123 Catania, Italy}

\corres{Correspondence: adelpopolo@oact.inaf.it}

\abstract{Motivated by the back-reaction debate, and~some unexplained characteristics
of the CMB, we investigate the possibility of some anisotropy in the
universe observed around us. To this aim, we build up a 
novel  
prediction for the Hubble law for the late universe from a Bianchi
type I model, 
{taken as proof of concept,}
 transcribing the
departure of such model from a $\Lambda$CDM model. We~dicussed the
redshift measurement in this universe, and~finally formalized the
Hubble diagram.}


\begin{document}

\section{Introduction}

One of the main assumption of the $\Lambda$CDM model is that, on large
scales, an isotropic and homogeneous spacetime can describe accurately
the universe, at least at the background level. While most of observations
agree with this assumptions, they show a clumpy matter distribution
on small scales. This point is at the core of a debate concerning
the magnitude of backreaction of the large scale structure on the
background dynamics. Anisotropy and~inhomogeneity might either explain
cosmic acceleration~\cite{Rasanen:2003fy,Rasanen:2006kp,Marozzi:2012ib},
or, according to other authors~\cite{Flanagan:2005dk}, have negligible
effects. Apart from this point, anisotropies in cosmological expansion
have been discussed since the early works of~\cite{Misner:1967uu,Gibbons:1977mu},
although their early models were favoring the late smearing of such
departures from isotropy. Together with the result that fundamental
observers measuring isotropic Cosmic Microwave Background \cite{Penzias:1965wn,Dicke:1965zz} [CMB]
radiation implied, we are in a spatially almost homogeneous and isotropic
region \cite{Stoeger:1999tb,Maartens:1994qq,Maartens:1995zz}, this
favored the dominant idea that the universe is an almost Friedman--Lemaître--Robertson--Walker
\cite{Friedman:1922kd,Lemaitre:1931zz,Robertson:1933zz,Walker:1933} [FLRW]
spacetime over keeping a fading out anisotropic behavior, as in Ref.~\citep{Collins:1973lda}. 

However, since the measurements of the CMB anisotropies in COBE~\cite{Mather:1991pc},
WMAP~\cite{Bennett:2003ba} and Planck~\cite{PlanckHFICoreTeam:2011az}
satellites, hints of a power hemispherical asymmetry have been found
and studied since the results of WMAP~\cite{Eriksen:2003db,Hansen:2004vq,Jaffe:2005pw,Hoftuft:2009rq},
continuing on Planck~\cite{Ade:2013nlj,Akrami:2014eta}, revealing
a ``preferred axis''~\cite{Zhao:2017unp} for the low angular resolution
part of the radiation temperature spectrum and its polarization~\cite{Contreras:2017qkc,ODwyer:2019rfg},
a~quadrupole--octupole alignment~\cite{Schwarz:2004gk,Copi:2005ff,Copi:2006tu,Copi:2010na,Copi:2013jna,Marcos-Caballero:2019jqj}
and a cold spot~\cite{Cruz:2004ce,Cruz:2006fy,Cruz:2006sv}. Moreover,
a recent emergence of a tension in the Hubble parameter measurements
\citep{Bolejko:2017fos} between the values inferred from large scale
expansion of the Planck measured CMB~\cite{Aghanim:2018eyx} $H_{0}=67.4\pm0.5$
km/s/Mpc at 68\% C.L., assuming the standard $\Lambda$CDM model,
and those obtained from the more local SNIa measurements~\cite{Riess:2018uxu,Riess:2018byc,Riess:2019cxk},
the~latest yielding $H_{0}=74.03\pm1.42$ km/s/Mpc at 68\% C.L. Those
anomalies seem to persist, although statistical effects for the
CMB~\cite{Bennett:2010jb}, and~systematic errors in SNIa have been
invoked~\cite{Rameez:2019wdt}.

Interrogations are thus piling up to reopen the case for anisotropic
expansion in the local universe. Bianchi models have been proposed,
escaping the prescriptions of Ref.~\citep{Collins:1973lda}, to explain
the CMB anomaly~\cite{Pontzen:2007ii,Sung:2010yv,Russell:2013oda}.
A recent PhD thesis was even produced, discussing anisotropic universes
\citep{Macpherson:2019cdo}, and~other types of non-FLRW models are
being explored~\cite{Cea:2019gnu}. 

Our study, which aims at producing a Hubble law for the late universe
from a Bianchi type I model, and~is amply motivated by all the above
hints, is therefore very timely and would be eminently useful in future
direct confrontation with observations of SNIa. {Since this is
a proof of concept study, we choose to investigate the simplest anisotropic
model away from the FLRW model that is with one anisotropy direction.
The novelty of the work resides in the construction of the confrontation
of this simple model with Hubble diagram observations.}

We organize the paper following Section~\ref{sec:Anisotropic-CDM-model},
describing the model we used to transcribe the Bianchi I model into
an appearant almost $\Lambda$CDM model. 
Section~\ref{sec:Redshift-in-anisotropic}
discuss the measurement of redshifts in such universe, while Section~\ref{sec:Hubble-law-in}
formalizes the Hubble diagram expressed in this model. Finally, Section~\ref{sec:Preliminary-confrontation-with}
proposes a preliminary test of the model, before concluding in Section~\ref{sec:Conclusions}.

\section{Anisotropic $ \Lambda$CDM Model \label{sec:Anisotropic-CDM-model}}

Our aim is to propose a model capable of including a level of anisotropies
compatible with observations that is looking very much like a $\Lambda$CDM
model in the past and developing anisotropies into the present. To
do so, we propose an expression of a Bianchi I model in a form similar
to the FLRW solution, and we develop its solution in order to keep
as close as possible to the derivations of the~FLRW.

\subsection{Model Setup}

We want to investigate the simplest anisotropic model away from the
isotropic and homogeneous FLRW model. We~are thus led to concentrate
on a flat Bianchi type I metric~\cite{Jacobs:1968zz} which we choose
in the form 
\begin{align}
ds^{2}= & -dt^{2}+a^{2}(t)\left[\left(1+\epsilon(t)\right)^{2}dx^{2}+dy^{2}+dz^{2}\right],\label{eq:metric}
\end{align}
where $a$ gives a global scale factor and we have chosen one direction
to expand anisotropically from the others, for which this departure
from isotropy is measured by $\epsilon$, the~anisotropic perturbation
parameter. {It is a measure of the anisotropic departure of the
model from the flat FLRW model by only multiplying the FLRW scale
factor $a(t)$ by an amount $\left(1+\epsilon(t)\right)$ in the chosen
$x$ direction of anisotropy. }The Einstein's field equations, in
this metric, for a perfect fluid of energy density $\rho$ and pressure
$p$ give 
\begin{align}
G_{\:t}^{t}=3\left(\frac{\dot{a}}{a}\right)^{2}+2\frac{\dot{a}}{a}\frac{\dot{\epsilon}}{\left(1+\epsilon\right)}= & \kappa\rho+\Lambda,\\
G_{\:y}^{y}=G_{\:z}^{z}=-2\frac{\ddot{a}}{a}-\left(\frac{\dot{a}}{a}\right)^{2}= & \kappa p-\Lambda,\\
G_{\:x}^{x}=-2\frac{\ddot{a}}{a}-\left(\frac{\dot{a}}{a}\right)^{2}-\frac{\ddot{\epsilon}}{1+\epsilon}-3\frac{\dot{a}}{a}\frac{\dot{\epsilon}}{1+\epsilon}= & \kappa p-\Lambda,
\end{align}
while the Bianchi identity yields 
\begin{align}
\dot{\rho}+\left(3\frac{\dot{a}}{a}+\frac{\dot{\epsilon}}{\left(1+\epsilon\right)}\right)\left(\rho+p\right)= & 0.
\end{align}
We further restrict to $\Lambda$CDM a dust fluid with a cosmological
constant, for which the pressure equations become purely geometrical,
\begin{align}
2\frac{\ddot{a}}{a}+\left(\frac{\dot{a}}{a}\right)^{2}+\frac{\ddot{\epsilon}}{1+\epsilon}+3\frac{\dot{a}}{a}\frac{\dot{\epsilon}}{1+\epsilon}= & \Lambda,\label{eq:AnisoAccPert}\\
2\frac{\ddot{a}}{a}+\left(\frac{\dot{a}}{a}\right)^{2}= & \Lambda,\label{eq:isoAcc}
\end{align}
while the remaining inhomogeneous set remains as 
\begin{align}
3\left(\frac{\dot{a}}{a}\right)^{2}+2\frac{\dot{a}}{a}\frac{\dot{\epsilon}}{\left(1+\epsilon\right)}= & \kappa\rho+\Lambda,\label{eq:AnisoFriedman}\\
\dot{\rho}+\left(3\frac{\dot{a}}{a}+\frac{\dot{\epsilon}}{\left(1+\epsilon\right)}\right)\rho= & 0,\label{eq:BianchiId}
\end{align}
The Friedmann-like Equation~\eqref{eq:isoAcc} can be rewritten in 
\begin{align*}
\left(a\dot{a}^{2}\right)^{\cdot}= & \frac{\Lambda}{3}\left(a^{3}\right)^{\cdot},
\end{align*}
to integrate into 
\begin{align}
\dot{a}^{2}= & \frac{a_{0}^{3}H_{0}^{2}\Omega_{0}}{a}+\frac{\Lambda}{3}a^{2}=H_{0}^{2}\left(\frac{a_{0}^{3}\Omega_{0}}{a}+\Omega_{\Lambda}a^{2}\right),\label{eq:AnisoHubble}
\end{align}
where the constant of integration is written with $a_{0}$, the~scale
factor with the $0$ index denoting values taken nowadays, $H_{0}=\frac{\dot{a}_{0}}{a_{0}}$,
the Hubble parameter at present, and~an arbitrary parameter $\Omega_{0}$,
a relative energy density to critical, and~where we have used the
definition $\Lambda=3H_{0}^{2}\Omega_{\Lambda}$. The~Bianchi identity
(\ref{eq:BianchiId}) integrates into 
\begin{align}
\rho= & \rho_{0}\left(\frac{a_{0}}{a}\right)^{3}\left(\frac{1+\epsilon_{0}}{1+\epsilon}\right),\label{eq:AnisoDen}
\end{align}
and we can combine (\ref{eq:isoAcc}) with (\ref{eq:AnisoAccPert})
to get the anisotropic perturbation parameter derivative: 
\begin{align}
\frac{\ddot{\epsilon}}{1+\epsilon}+3\frac{\dot{a}}{a}\frac{\dot{\epsilon}}{1+\epsilon}= & 0.
\end{align}
We then assume $\epsilon$ to represent a vanishing perturbation at
some initial time just before the recombination ($\epsilon\underset{t\to t_{i}}{\longrightarrow}0$,
setting $\frac{a_{r}}{a_{i}}=10^{n}$, with $\epsilon_{r}\sim10^{-5}$
at $\frac{a_{r}}{a_{0}}=10^{-3}$~\cite{Smoot:1992td,PlanckHFICoreTeam:2011az,Ade:2013nlj}{)
Here, we use index $i$ to designate values at this initial time and
index $r$ to indicate values at recombination. The~recombination
scale ratio and magnitude of fluctuations are extracted from the references,
and we assume that fluctuations caused by anisotropy imply such anisotropy
to be of the same magnitude. We~use the power $n$ of 10 to mark the
scale growth between the initial and recombination epochs. Its value
will be determined from the fit of the model to the data. We} 
 can rearrange and integrate the derivative equation into 
\begin{align}
\dot{\epsilon}= & \dot{\epsilon}_{0}\left(\frac{a_{0}}{a}\right)^{3},\label{eq:deltaDot}
\end{align}
with the constant of integration reformulated such as to appear with
$\dot{\epsilon}_{0}$, the~anisotropic perturbation parameter derivative
at present. Combining it with the Bianchi identity solution (\ref{eq:AnisoDen})
allows, from (\ref{eq:AnisoFriedman}), for finding the scale evolution
equation 

\begin{align}
3\left(\frac{\dot{a}}{a}\right)^{2}+2\frac{\dot{a}}{a^{4}}\frac{\dot{\epsilon}_{0}a_{0}^{3}}{1+\epsilon}= & \kappa\rho_{0}\left(\frac{a_{0}}{a}\right)^{3}\left(\frac{1+\epsilon_{0}}{1+\epsilon}\right)+\Lambda.\label{eq:aEvol}
\end{align}

\subsection{Anisotropy-Scale Relation Interpretation}

From Equation~\eqref{eq:AnisoExact} of Appendix \ref{part:Appendix-A:-Solutions},
we can obtain an exact form for $\epsilon$ such that the constraints
in the CMB~\cite{Smoot:1992td,PlanckHFICoreTeam:2011az,Ade:2013nlj}
lead to a current order of magnitude expected for the model: since $\frac{a_{r}}{a_{0}}=10^{-3}\ll1$
so $\frac{a_{i}}{a_{0}}=10^{-\left(n+3\right)}\ll1$ ($n$ is assumed
around 2) and the $\Omega$s are of order one,
\begin{align}
1-\frac{\epsilon_{r}}{\epsilon_{0}}= & \frac{\sqrt{\Omega_{0}\left(\frac{a_{r}}{a_{0}}\right)^{-3}+\Omega_{\Lambda}}-1}{\sqrt{\Omega_{0}\left(\frac{a_{i}}{a_{0}}\right)^{-3}+\Omega_{\Lambda}}-1}\nonumber \\
\simeq & \left(\frac{a_{r}}{a_{i}}\right)^{-\frac{3}{2}}=10^{-\frac{3}{2}n}\\
\Rightarrow\epsilon_{0}\simeq & \epsilon_{r}\left(1+10^{-\frac{3}{2}n}\right)\simeq10^{-5},
\end{align}
so the evolution of the anisotropy is very small nowadays compared
to recombination.

\section{Redshift in Anisotropic Models \label{sec:Redshift-in-anisotropic}}

Since we want to produce a Hubble diagram for our model that is a
Hubble parameter evolution with redshift capable of confrontation
with the observable redshift vs magnitude relationship, we need to
make the connection of our model with the measurements of redshifts explicit.

In our anisotropic model, the~redshift $z$ depends on the direction
of observation: along any direction orthogonal to the $x$-axis, we
can expect, by analogy with FLRW, that $z=\frac{1}{a}-1$, while, along
the $x$ axis, we can expect $z=\frac{1}{a\left[1+\epsilon\left(a\right)\right]}-1$.
In general, we need to define the comoving distance in any given direction
to measure real distances as we expect the redshift to be defined
in terms of observed and emitted wavelengths $\lambda_{o}$ and $\lambda_{e}$
by the form $z=\frac{\lambda_{o}}{\lambda_{e}}-1$.

\subsection{Comoving Distance}

We place the observer at the center of the frame and measure the comoving
distance in an arbitrary direction. We~start from the line element
of lightlike trajectories proceeding from Equation~\eqref{eq:metric}
for $ds=0$ 
\begin{align}
1= & a^{2}(t)\left[\left(1+\epsilon(t)\right)^{2}\left(\frac{dx}{dt}\right)^{2}+\left(\frac{dy}{dt}\right)^{2}+\left(\frac{dz}{dt}\right)^{2}\right].
\end{align}
We then define $l$, the~projected comoving coordinate orthogonal
to the $x$-axis and $R$, the~comoving distance, as well as the angle
$\alpha$ of the observed comoving source with respect to the $x$-axis, with 
\begin{align}
dl^{2}= & dy^{2}+dz^{2},\\
dR^{2}= & dx^{2}+dl^{2},\\
\frac{dl}{dx}= & \tan\alpha.
\end{align}
In these terms, the~comoving distance obeys 
\begin{align}
1= & a^{2}\left(\left(1+\epsilon\right)^{2}\cos^{2}\alpha+\sin^{2}\alpha\right)\left(\frac{dR}{dt}\right)^{2}
\end{align}
which integrates, between emission and reception times $t_{e}$ and
$t_{o}$, into 
\begin{align}
R= & \int_{t_{e}}^{t_{o}}\frac{dt}{a\sqrt{\left(1+\epsilon\right)^{2}\cos^{2}\alpha+\sin^{2}\alpha}}.\label{eq:ComDist}
\end{align}
This allows us to calculate the redshift in this anisotropic framework.

\subsection{Redshift Calculation}

The comoving distance covered by light emitted at $t_{e}$ from $R_{e}$
and received at $t_{o}$ by an observer at $R_{o}=0$ is the same
as that emitted after one period at $t_{e}+\delta t_{e}$from $R_{e}$
and received at $t_{o}+\delta t_{o}$ by an observer at $R_{o}=0$.
From Equation~\eqref{eq:ComDist}, we express the previous statement as:
\begin{align}
\int_{t_{e}}^{t_{o}}\frac{dt}{a\sqrt{\left(1+\epsilon\right)^{2}\cos^{2}\alpha+\sin^{2}\alpha}}=
\int_{t_{e}+\delta t_{e}}^{t_{o}+\delta t_{o}}\frac{dt}{a\sqrt{\left(1+\epsilon\right)^{2}\cos^{2}\alpha+\sin^{2}\alpha}},
\end{align}
which is equivalent, from the properties of integrals, to equating
the comoving distances covered by light in one period (comoving wavelengths)
at emitter and observer 
\begin{align}
\int_{t_{e}}^{t_{e}+\delta t_{e}}\frac{dt}{a\sqrt{\left(1+\epsilon\right)^{2}\cos^{2}\alpha+\sin^{2}\alpha}}=
\int_{t_{0}}^{t_{0}+\delta t_{0}}\frac{dt}{a\sqrt{\left(1+\epsilon\right)^{2}\cos^{2}\alpha+\sin^{2}\alpha}}.
\end{align}
For any wavelength much shorter than the distance to the source (recall
$c=1$), $\lambda=\delta t\ll t$, we can consider that, over such
interval of time, the values of the scale factors are constant, so
the integrals above can be approximated by 
\begin{align}
\frac{\delta t_{e}}{a_{e}\sqrt{\left(1+\epsilon_{e}\right)^{2}\cos^{2}\alpha+\sin^{2}\alpha}}\approx
\frac{\delta t_{o}}{a_{o}\sqrt{\left(1+\epsilon_{o}\right)^{2}\cos^{2}\alpha+\sin^{2}\alpha}}.
\end{align}
Dropping the emitter's index and observing nowadays, we get 
\begin{align}
z\equiv\frac{\lambda_{0}}{\lambda}-1\approx & \frac{a_{0}}{a}\sqrt{\frac{1+\left[2+\epsilon_{0}\right]\epsilon_{0}\cos^{2}\alpha}{1+\left[2+\epsilon\right]\epsilon\cos^{2}\alpha}}-1.\label{eq:ExactMuRedshiftRel}
\end{align}
Further restricting to linear order in $\epsilon$, we obtain the linearized
relation between redshift, direction, and both scale factors 
\begin{align}
1+z= & \frac{a_{0}}{a}\sqrt{\frac{1+2\epsilon_{0}\cos^{2}\alpha}{1+2\epsilon\cos^{2}\alpha}}.\label{eq:LinDeltaRedshiftRel}
\end{align}

\subsection{Angle Averaging and Scale-Small Anisotropic Deviation Redshift Relation}

As observations are not generally taking into account the possibility
of a direction dependent expansion, we need to produce a direction
independent evaluation of the impact of anisotropies. We~start from
the general Equation~\eqref{eq:ExactMuRedshiftRel} and proceed to average
over all angles: expressing the redshift-scale relation as 
\begin{align}
1+\left[2+\epsilon\right]\epsilon\cos^{2}\alpha=
\left(\frac{a_{0}}{a}\right)^{2}\frac{1}{\left(1+z\right)^{2}}\left(1+\left[2+\epsilon_{0}\right]\epsilon_{0}\cos^{2}\alpha\right),
\end{align}
averaging over all $\alpha$s produces factors of 1/2 
for each $\cos^{2}\alpha$ factor, so we get 
\begin{align}
\left(1+z\right)^{2}\overline{\left(\frac{a}{a_{0}}\right)^{2}}= & \frac{1+\left[1+\frac{\overline{\epsilon_{0}}}{2}\right]\overline{\epsilon_{0}}}{1+\left[1+\frac{\overline{\epsilon}}{2}\right]\overline{\epsilon}},\label{eq:exactAngleAveragedRedshift}
\end{align}
which, to a linear order, can be written as the angle-averaged, linearized,
redshift-scale relation 
\begin{align}
\left(1+z\right)^{2}\overline{\left(\frac{a}{a_{0}}\right)^{2}}\simeq & 1+\left(\overline{\epsilon_{0}}-\overline{\epsilon}\right)=1+\overline{\epsilon_{0}}\left(1-\frac{\overline{\epsilon}}{\overline{\epsilon_{0}}}\right).\label{eq:AngleAveragedZ}
\end{align}
Note that we recover the isotropic $z=0$ at the present time.

\section{Hubble Law in Anisotropic Models \label{sec:Hubble-law-in}}

The usual FLRW model produces a Hubble law by computing the Hubble
parameter evolution as a function of redshift. In order to easily
confront our model with the isotropic standard, we need to express
in the framework of our Bianchi I anisotropically expanding model
a similarly formulated Hubble law.

\subsection{Generalized Hubble Parameter}

Now, we want the Friedmann-equivalent Hubble parameter to confront
FLRW-based measurements done in an anisotropic universe. We~can define
the Hubble parameter as the rate of relative volume change. In FLRW
models, such rate is related to the expansion scalar and to the FLRW
Hubble parameter straightforwardly 
\begin{align}
\frac{\dot{V}}{3V}= & \frac{\Theta}{3}=H_{FLRW}.
\end{align}
Assuming the FLRW model, Hubble measurements averaged with the angle give
access to the expansion rate. In this anisotropic model, the~expansion
rate can be found in the Bianchi identity, which can be recast~as
\begin{align}
\frac{d}{dt}\ln\rho= & -\Theta=-\left(3\frac{\dot{a}}{a}+\frac{\dot{\epsilon}}{1+\epsilon}\right).\label{eq:Expansion}
\end{align}
Thus, defining the FLRW-like angle averaged Hubble parameter from the
expansion $\mathcal{H}=\frac{d\ln V}{3dt}=\frac{1}{3}\Theta$, we
get from Equation~\eqref{eq:Expansion} 
\begin{align}
\mathcal{H}= & \frac{\dot{a}}{a}+\frac{\dot{\epsilon}}{3\left(1+\epsilon\right)}.\label{eq:HubbleAnioDef}
\end{align}
Solving Equation~\eqref{eq:AnisoFriedman} for $\frac{\dot{a}}{a}=H$,
and selecting the positive root (recall from (\ref{eq:AnisoDen})
$\kappa\rho+\Lambda=3H_{0}^{2}\left(\Omega_{m}\left(\frac{a_{0}}{a}\right)^{3}\left(\frac{1+\epsilon_{0}}{1+\epsilon}\right)+\Omega_{\Lambda}\right)$,
with $3H_{0}^{2}\Omega_{m}=\kappa\rho_{0}$), we get 
\begin{align}
\frac{\dot{a}}{a}= & \sqrt{\left(\frac{\dot{\epsilon}}{3\left(1+\epsilon\right)}\right)^{2}+H_{0}^{2}\left(\Omega_{m}\left(\frac{a_{0}}{a}\right)^{3}\left(\frac{1+\epsilon_{0}}{1+\epsilon}\right)+\Omega_{\Lambda}\right)}\nonumber \\
 & -\frac{\dot{\epsilon}}{3\left(1+\epsilon\right)}\label{eq:dlnaSdt}
\end{align}
From (\ref{eq:AnisoHubble}) taken now, we can write 
\begin{align*}
1= & \Omega_{0}+\Omega_{\Lambda},
\end{align*}
while the same treatment applied to Equation~\eqref{eq:AnisoFriedman}
yields 
\begin{align*}
1+\frac{2}{3H_{0}}\frac{\dot{\epsilon}_{0}}{1+\epsilon_{0}}= & \Omega_{m}+\Omega_{\Lambda}.
\end{align*}
The two previous relations yield 
\begin{align}
\Omega_{\Lambda}= & 1-\Omega_{0}=1-\Omega_{m}+\frac{2}{3H_{0}}\frac{\dot{\epsilon}_{0}}{1+\epsilon_{0}}.
\end{align}
We can thus rewrite, from Equations~\eqref{eq:HubbleAnioDef} and \eqref{eq:dlnaSdt}
and the previous expression for $\Omega_{\Lambda}$, the~Hubble parameter
into 
\begin{align}
\mathcal{H}^{2}= & H_{0}^{2}\left(\Omega_{m}\left[\left(\frac{a_{0}}{a}\right)^{3}\left(\frac{1+\epsilon_{0}}{1+\epsilon}\right)-1\right]+1+\frac{2}{3H_{0}}\frac{\dot{\epsilon}_{0}}{1+\epsilon_{0}}\right).\label{eq:Hubble(delta,a)=00003D00003DH(z)}
\end{align}

\subsection{Hubble-Scale-Redshift Relation}

At this point, we shall use Equations~\eqref{eq:deltaDot}
and \eqref{eq:exactAngleAveragedRedshift} to introduce the redshift and anisotropic
perturbation parameter's derivative. We~first rewrite Equation~\eqref{eq:exactAngleAveragedRedshift} into 
\begin{align}
\frac{a_{0}}{a}= & \left(1+z\right)\sqrt{\frac{1+\left[1+\frac{\epsilon}{2}\right]\epsilon}{1+\left[1+\frac{\epsilon_{0}}{2}\right]\epsilon_{0}}},
\end{align}
so the factors involved in Equation~\eqref{eq:Hubble(delta,a)=00003D00003DH(z)},
in terms of redshift, linearized in $\epsilon$ read (Equations~\eqref{eq:ddelta/1+delta}--\eqref{eq:denScaling}
of Appendix \ref{part:Appendix-B:-Anisotropic}) 
\begin{align}
\frac{\dot{\epsilon}}{3\left(1+\epsilon\right)}\simeq & \frac{\dot{\epsilon}_{0}}{3}\left(1+z\right)^{3}\left(1+\frac{\epsilon_{0}}{2}\left[\frac{\epsilon}{\epsilon_{0}}-3\right]\right),\\
\left(\frac{a}{a_{0}}\right)^{3}\simeq & \left(1+z\right)^{-3}\left(1-\frac{3}{2}\epsilon_{0}\left[\frac{\epsilon}{\epsilon_{0}}-1\right]\right),\\
\left(\frac{a_{0}}{a}\right)^{3}\frac{1+\epsilon_{0}}{1+\epsilon}\simeq & \left(1+z\right)^{3}\left(1+\frac{\epsilon_{0}}{2}\left[\frac{\epsilon}{\epsilon_{0}}-1\right]\right).
\end{align}
The Hubble parameter from Equation~\eqref{eq:Hubble(delta,a)=00003D00003DH(z)}
then reads, in terms of the above redshift expressions, the~derivative
in Equation~\eqref{eq:dd0/H0} from Appendix \ref{part:Appendix-A:-Solutions}
and the definition (\ref{eq:AnisoDepSolLin}) from Appendix \ref{part:Appendix-B:-Anisotropic}

\begin{align}
\mathcal{H}\simeq & H_{0}\sqrt{\Omega_{m}\left[\left(1+z\right)^{3}\left(1+\frac{\epsilon_{0}}{2}\left[\frac{\epsilon}{\epsilon_{0}}-1\right]\right)-1\right]+1+\frac{2}{3}\frac{\epsilon_{0}}{\Delta_{0}}}\nonumber \\
 & \textrm{with }\Delta_{0}=\frac{2}{\Omega_{m}}\left[\sqrt{\Omega_{m}\left(\left(1+z_{i}\right)^{3}-1\right)+1}-1\right].
\end{align}

\subsection{Hubble-Redshift Relation}

From Equation~\eqref{eq:AnisoDepSolLin} of Appendix \ref{part:Appendix-B:-Anisotropic},
we can further obtain an integral form for $\epsilon$: 
\begin{align}
\epsilon\simeq & \epsilon_{0}\frac{\left[\sqrt{\Omega_{m}\left(\left(1+z_{i}\right)^{3}-1\right)+1}-\sqrt{\Omega_{m}\left(\left(1+z\right)^{3}-1\right)+1}\right]}{\left[\sqrt{\Omega_{m}\left(\left(1+z_{i}\right)^{3}-1\right)+1}-1\right]}.
\end{align}

We now have the tools to get the Hubble redshift relation, simplifying
the expression (\ref{eq:H(z)}) of Appendix \ref{part:Appendix-B:-Anisotropic},
in a shape close to the FLRW form

\begin{align}
I\left(z\right)= & \sqrt{\Omega_{m}\left(\left(1+z\right)^{3}-1\right)+1},\\
I\left(z_{r}\right)= & \sqrt{\Omega_{m}\left(10^{9}-1\right)+1},\\
I_{0}= & \sqrt{\Omega_{m}\left(10^{3\left(n+3\right)}-1\right)+1},\\
\mathcal{H}\left(z\right)\simeq & H_{0}\sqrt{\Omega_{m}\left(\left(1+z\right)^{3}\left\{ 1-\frac{\epsilon_{r}}{2}\frac{1-I\left(z\right)}{I_{0}-I\left(z_{r}\right)}\right\} -1\right)}\nonumber \\
 & \overline{+\left\{ 1+\frac{\Omega_{m}}{3}\frac{\epsilon_{r}}{I_{0}-I\left(z_{r}\right)}\right\} }\label{eq:modelGeneral}\\
\simeq & H_{0}\sqrt{\Omega_{m}\left(1+z\right)^{3}\left\{ 1-\frac{\epsilon_{r}}{2}\frac{1-I\left(z\right)}{I_{0}-I\left(z_{r}\right)}\right\} }\nonumber \\
 & \overline{+\Omega_{\Lambda}}\label{eq:model}\\
\Omega_{\Lambda}= & 1-\Omega_{m}\left(1-\frac{1}{3}\frac{\epsilon_{r}}{I_{0}-I\left(z_{r}\right)}\right).
\end{align}

{This} concludes the results of our model.

\section{Preliminary Confrontation with Observations \label{sec:Preliminary-confrontation-with}}

\begin{figure}[H]
\includegraphics[width=15cm]{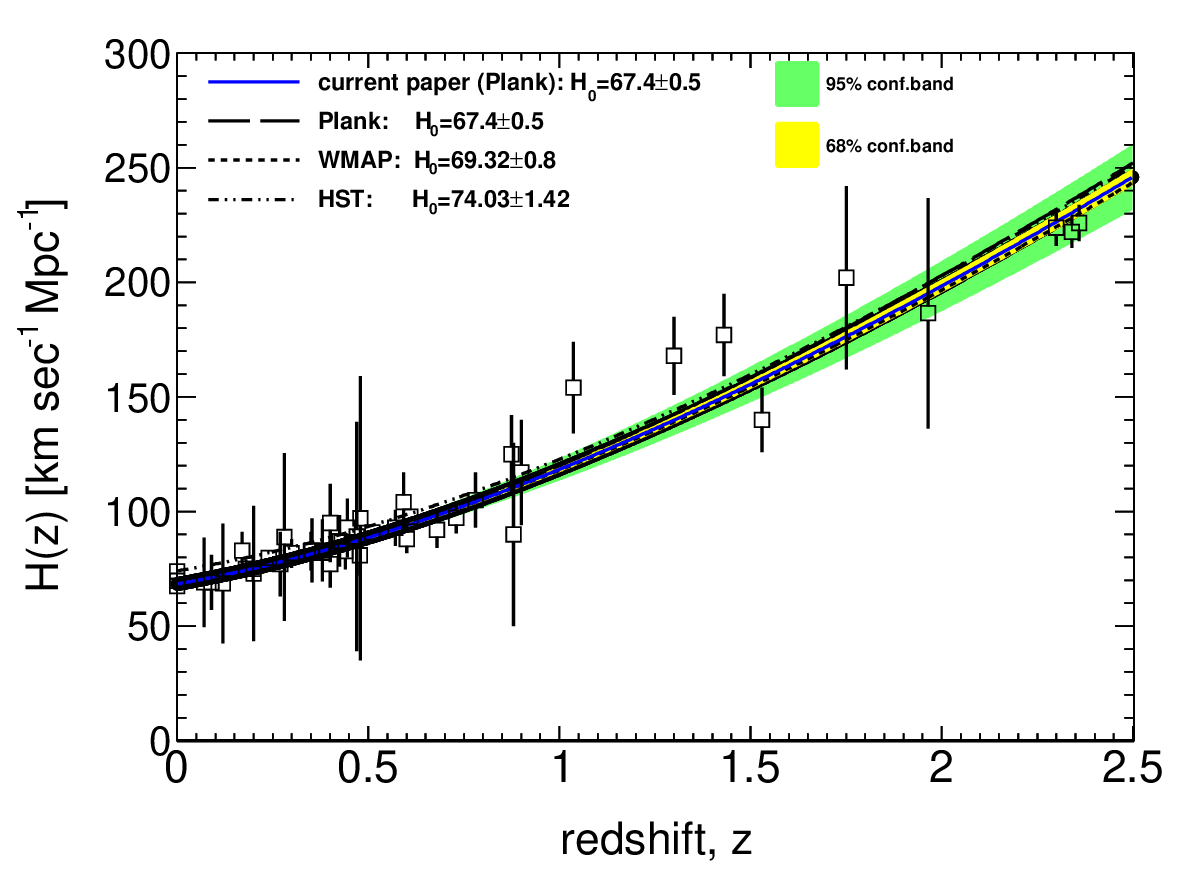}
\caption{Hubble diagram confrontation
of $\Lambda$CDM and our model Equation~\eqref{eq:model} assuming that
$H_{0}$ was taken from the Planck, WMAP, and HST results.}
\label{fig:Hubble-diagram-confrontation}
\end{figure}

We present here a first approach to using the model in detection of
anisotropy. We~obtained Hubble evolution from {the Gemini Deep
Survey (GDDS), Sloan Digital Sky Survey III Baryon Oscillation Spectroscopic
Survey (SDSS-III), and~highest redshift Ly$\alpha$ measurements.
These data samples provide high-quality spectroscopy of red galaxies,
some of which show stellar absorption features, indicating an old
stellar population. The~differential aging of these cosmic chronometers
has been used to measure the observed Hubble parameter at different
redshifts reported in Refs.~\citep{Moresco:2012jh,Anderson:2013oza,Simon:2004tf,Yu:2017iju},
and does not include older results from earlier analyses of data subsets,
or estimates that are no longer trusted to be reliable. U}sing
the parameterised evolution
\begin{align*}
H(z)= & H_{0}\sqrt{\Omega_{m}\left(1+z\right)^{3}+\Omega_{\Lambda}},
\end{align*}
confronted with our model Equation~\eqref{eq:model}, for which we take
the $H_{0}$ value from the Planck, WMAP, or HST evaluations, and $\Omega_{m}$
and $\epsilon_{r}$ are free parameters. {This was done to study
the effect of the assumed $H_{0}$ value on our results. }By combining
data sets taken from~\cite{Simon:2004tf,Moresco:2012jh,Anderson:2013oza,Sharov:2014voa,Yu:2017iju},
a fit is done with the help of the Log-Likelihood and $\chi^{2}$
methods implemented in the ROOT framework~\cite{Antcheva:2009zz},
for which the results on the anisotropy parameter are summarized in Table~\ref{tab:Model-parameters-analysis}.
{The obtained $\Omega_{m}$ from the fits were consistent with
standard results and we did not estimate them worth indicating. }This
allows for plotting the Hubble diagrams shown in Figure~\ref{fig:Hubble-diagram-confrontation}.
The differences in assumptions of the $H_{0}$ parameter do not noticeably affect
 the results (the lines overlap in Figure~\ref{fig:Hubble-diagram-confrontation});
however, as can be seen in Table~\ref{tab:Model-parameters-analysis},
it does change the evaluation of the anisotropy parameter and its
variance $\epsilon_{r}\pm\sigma$, and~somehow so does the regression
method.

In Figure~\ref{fig:Hubble-diagram-confrontation}, the~black lines
(dashed, short-dashed, double dotted-dashed) represent the Hubble
diagrams for the $\Lambda$CDM model fits with the fixed $H_{0}$
parameter with respect to the recent reports of the experimental collaborations,
namely Plank~\cite{Aghanim:2018eyx}, WMAP~\cite{2011ApJS.192.18K}
and HST~\cite{Riess:2019cxk}, while $\Omega_{m}$ and $\epsilon_{r}$
were released. The~square data points with error bars are all taken
from~\cite{Simon:2004tf,Moresco:2012jh,Anderson:2013oza,Sharov:2014voa,Yu:2017iju}.
The colored line (solid blue) display our model using the $H_{0}$
determined from the Plank report~\cite{Aghanim:2018eyx}. The~light
yellow and green bands show its 1 and 2$\sigma$ confidence levels,
respectively. 

Our model matches within a 68\% confidence level to the reference $\Lambda$CDM
model, and {thus is} not significantly different from the isotropic
case. This is reflected in Table~\ref{tab:Model-parameters-analysis},
where the values of the anisotropy parameters, with their variances,
are compatible with isotropy, except for the WMAP choice of $H_{0}$
combined either with the Log-Likelihood method or with the $\chi^{2}$
and empty bins weighting method, where a slight detection is obtained.
{The weighting methods are recommended to be used in case of low
statistics and when data represent counts with Poisson statistics.
The $\chi^{2}$ method without empty bins weighing results in very
huge uncertainties for $\epsilon_{r}$. Except for the unweighted
method, the~large size of the variances should be noted. Therefore,
} clear detection would require cleaner data or larger sampling
of the redshift range.

\begin{table}[H]
\centering
\caption{Model parameters analysis to
obtain $\epsilon_{r}\pm\sigma$. We~fit the data from~\cite{Simon:2004tf,Moresco:2012jh,Anderson:2013oza,Sharov:2014voa,Yu:2017iju}
with two different methods, namely Log-Likelihood and $\chi^{2}$.
In addition to fit method, there are two weighting methods for the
fitted distribution. The~designation ``weighting empty bins'' means
that we assign weights equal to unity to all bins, including empty
bins, while the method without weighting does not use any assumption
about the data-points weights. In all fits, we fixed the $H_{0}$ value
and release $\Omega_{M}$ and $\epsilon_{r}$ parameters using Equation~\eqref{eq:model}.
The fixed $H_{0}$ refers to the reports from the different experimental
collaboration, namely Plank~\cite{Aghanim:2018eyx}, WMAP~\cite{2011ApJS.192.18K},
and HST~\cite{Riess:2019cxk}. }
\label{tab:Model-parameters-analysis}
\begin{tabular}{>{\raggedright}p{0.5\textwidth}cccccc}
\toprule
\hspace*{\fill}\textbf{Fit method}\hspace*{\fill} & \multicolumn{2}{c}{\textbf{Plank} } & \multicolumn{2}{c}{\textbf{WMAP}} & \multicolumn{2}{c}{\textbf{HST}}\tabularnewline
\hline 
 & \boldmath$\epsilon_{r}$ & \boldmath$\sigma$ & \boldmath$\epsilon_{r}$ & \boldmath$\sigma$ & \boldmath$\epsilon_{r}$ & \boldmath$\sigma$\tabularnewline
\midrule
Log-Likelihood (with/without weighting empty bins) & 0.25  & 0.37  & 0.13 & 0.12  & 0.25  & 0.37 \tabularnewline
$\chi^{2}$ (without weighting empty bins) & 0.13  & 1.32  & 0.25  & 4.03  & 0.13  & 1..31 \tabularnewline
$\chi^{2}$ (with weighting empty bins) & 0.25  & 0.37  & 0.13  & 0.12  & 0.25  & 0.37\\\bottomrule
\end{tabular}
\end{table}

Although the level of uncertainties does not allow us to report definite values for $\epsilon_{r}$ and these results are not giving
clearly detectable anisotropy, we argue that the method could be used
as a complement to the dipole/quadrupole approach from Ref.~\citep{Bolejko:2015gmk}. 

\section{Conclusions \label{sec:Conclusions}}

In this paper, we have constructed a Hubble law capable of being confronted
with observations that were designed for an homogeneous and isotropic
expanding universe, while allowing for global anisotropy in expansion.
Such anisotropy would appear in observations designed without taking
it into account as distortions of the redshift-distance behavior.
This tool is obtained from solving the Einstein Field Equations for
a Bianchi I model that is almost FLRW. For this model, the effect of anisotropy on redshift is obtained which is then synthesized for
isotropy \textendash{} assuming observation by an angle averaged redshift
expression. We~finally produce a Hubble function of redshift that
we propose can be used in future confrontation with observations.
Such a tool, we argue, is needed now as we have several 
reasons, starting from the back-reaction debate, and~going on with
some unexplained characteristics of the CMB that hint at the possibility
that the role of anisotropy in the universe is not trivial. The~model
built in 
this paper 
discussed 
an anisotropic universe model that could be checked against CMB, or~SNIa, predictions. As a preliminary example of how to use the model,
we~confronted it with three data sets and found no conclusive results.
However, we argue that, with a more thorough approach, this method
could be a useful complement to other approaches such as that of
a Bolejko~\cite{Bolejko:2015gmk} 
late 
model 
measurement 
diagram.
The model we build up will be checked in the next paper against cosmological
data. 

%
\vspace{6pt}

\authorcontributions{all authors contributed equally} 

\funding{Lanzhou University starting
fund and the Fundamental Research Funds for the Central Universities
(Grant No. lzujbky-2019-25).
}

\acknowledgments{MLeD acknowledges the financial support by the Lanzhou University starting
fund and the Fundamental Research Funds for the Central Universities
(Grant No. lzujbky-2019-25). M.D. thanks Xin Wu for his support
in this research during the COVID-19 quarantine measures.}

\conflictsofinterest{The authors declare no conflict of interest.}  

%

\appendixtitles{yes}
\appendixsections{multiple}
\appendix

\section{Solutions to Anisotropic Scale Parameters \label{part:Appendix-A:-Solutions}}
\vspace{-6pt}

\subsection{Solutions with $ \Omega_{0}$}

From (\ref{eq:deltaDot}), one can integrate the anisotropic parameter,
using (\ref{eq:AnisoHubble}) and the variable change \mbox{$X=\frac{a}{a_{0}}$,
into }
\begin{align}
\epsilon= & \frac{\dot{\epsilon}_{0}}{H_{0}\sqrt{\Omega_{0}}}\int_{\frac{a_{i}}{a_{0}}}^{\frac{a}{a_{0}}}\frac{dX}{X^{\frac{5}{2}}\sqrt{1+\frac{\Omega_{\Lambda}}{\Omega_{0}}X^{3}}}.\label{eq:AnisoDepSol}
\end{align}
We can further use the variable change 
\begin{align}
Z= & \frac{\Omega_{\Lambda}}{\Omega_{0}}X^{3} & \Rightarrow X= & \left(\frac{\Omega_{0}}{\Omega_{\Lambda}}Z\right)^{\frac{1}{3}},\\
 &  & \Rightarrow dX= & \frac{1}{3}\left(\frac{\Omega_{0}}{\Omega_{\Lambda}}\right)^{\frac{1}{3}}\frac{dZ}{Z^{\frac{2}{3}}},
\end{align}
to obtain in general 
\begin{align}
\epsilon= & \frac{\dot{\epsilon}_{0}\sqrt{\Omega_{\Lambda}}}{3H_{0}\Omega_{0}}\left(\int_{\frac{\Omega_{\Lambda}}{\Omega_{0}}\left(\frac{a_{i}}{a_{0}}\right)^{3}}^{\frac{\Omega_{\Lambda}}{\Omega_{0}}\left(\frac{a}{a_{0}}\right)^{3}}Z^{-\frac{1}{2}-1}\left(1+Z\right)^{\frac{1}{2}-1}dZ\right)\nonumber \\
= & \frac{\dot{\epsilon}_{0}\sqrt{\Omega_{\Lambda}}}{3H_{0}\Omega_{0}}\left[-2\sqrt{1+\frac{1}{Z}}\right]_{\frac{\Omega_{\Lambda}}{\Omega_{0}}\left(\frac{a_{i}}{a_{0}}\right)^{3}}^{\frac{\Omega_{\Lambda}}{\Omega_{0}}\left(\frac{a}{a_{0}}\right)^{3}}\nonumber \\
= & \frac{2\dot{\epsilon}_{0}}{3H_{0}\Omega_{0}}\left[\sqrt{\Omega_{0}\left(\frac{a_{i}}{a_{0}}\right)^{-3}+\Omega_{\Lambda}}-\sqrt{\Omega_{0}\left(\frac{a}{a_{0}}\right)^{-3}+\Omega_{\Lambda}}\right],\label{eq:AnisoDepSolNL}
\end{align}
so we can build the ratio ($\Omega_{\Lambda}=1-\Omega_{0}$) 
\begin{align}
\frac{\epsilon_{r}}{\epsilon_{0}}= & \frac{\left[\sqrt{\Omega_{0}\left(\frac{a_{i}}{a_{0}}\right)^{-3}+\Omega_{\Lambda}}-\sqrt{\Omega_{0}\left(\frac{a_{r}}{a_{0}}\right)^{-3}+\Omega_{\Lambda}}\right]}{\left[\sqrt{\Omega_{0}\left(\frac{a_{i}}{a_{0}}\right)^{-3}+\Omega_{\Lambda}}-1\right]}.\label{eq:AnisoExact}
\end{align}

\subsection{Solutions with $ \Omega_{m}$}

Using $\Omega_{\Lambda}=1-\Omega_{0}$ and defining $x$ with $\Omega_{0}=\Omega_{m}-\frac{2}{3H_{0}}\frac{\dot{\epsilon}_{0}}{1+\epsilon_{0}}=\Omega_{m}-x$,
so $\frac{\dot{\epsilon}_{0}}{H_{0}}=\frac{3}{2}\left(1+\epsilon_{0}\right)x$,
we can rewrite Equation~\eqref{eq:AnisoDepSolNL}, at present time 
\begin{align}
\epsilon_{0}= & 3\frac{\left(1+\epsilon_{0}\right)x}{\left(\Omega_{m}-x\right)}\left[\sqrt{\left(1-\Omega_{m}+x\right)+\left(\Omega_{m}-x\right)\left(\frac{a_{i}}{a_{0}}\right)^{-3}}-1\right].
\end{align}
Assuming that $x\ll\Omega_{m}$, we linearize the above expression
in $x$. We~first linearize all factors 
\begin{multline*}
\epsilon_{0}\simeq3\frac{\left(1+\epsilon_{0}\right)}{\Omega_{m}}x\left[\sqrt{\Omega_{m}\left(\left(\frac{a_{i}}{a_{0}}\right)^{-3}-1\right)+1}-1\right]\\
\times\left(1+\left[\frac{1}{\Omega_{m}}-\frac{\left(\left(\frac{a_{i}}{a_{0}}\right)^{-3}-1\right)}{2\left(\Omega_{m}\left(\left(\frac{a_{i}}{a_{0}}\right)^{-3}-1\right)+1-\sqrt{\Omega_{m}\left(\left(\frac{a_{i}}{a_{0}}\right)^{-3}-1\right)+1}\right)}\right]x+O(x^{2})\right),
\end{multline*}
to finally simplify the linearization and retain only the linear terms
\begin{align}
\epsilon_{0}\simeq & 3\frac{\left(1+\epsilon_{0}\right)}{\Omega_{m}}\left[\sqrt{\Omega_{m}\left(\left(\frac{a_{i}}{a_{0}}\right)^{-3}-1\right)+1}-1\right]x.
\end{align}
Finally, solving for $x$, we find the expression of the derivative
in the constant of integration in terms of the present anisotropic
perturbation parameter and define the constant $\Delta_{0}$ 
\begin{align}
\frac{\dot{\epsilon}_{0}}{H_{0}}= & \frac{3}{2}\left(1+\epsilon_{0}\right)x\nonumber \\
\simeq & {\epsilon_{0}}/{\frac{2}{\Omega_{m}}\left[\sqrt{\Omega_{m}\left(\left(\frac{a_{i}}{a_{0}}\right)^{-3}-1\right)+1}-1\right]}\nonumber \\
= & \frac{\epsilon_{0}}{\Delta_{0}}\label{eq:dd0/H0}\\
\textrm{with }\Delta_{0}= & \frac{2}{\Omega_{m}}\left[\sqrt{\Omega_{m}\left(\left(\frac{a_{i}}{a_{0}}\right)^{-3}-1\right)+1}-1\right].
\end{align}
\vspace{-0.5cm}

\section{Anisotropic Redshift Hubble Calculations \label{part:Appendix-B:-Anisotropic}}
\vspace{-6pt}

\subsection{Anisotropic Redshift Terms}

Applying Equation~\eqref{eq:exactAngleAveragedRedshift} to introduce
the redshift relation to the scale factors, we have the following:
\begin{align}
\frac{a_{0}}{a}= & \left(1+z\right)\sqrt{\frac{1+\left[1+\frac{\epsilon}{2}\right]\epsilon}{1+\left[1+\frac{\epsilon_{0}}{2}\right]\epsilon_{0}}}.
\end{align}
From above, we can also compose 
\begin{align}
\left(\frac{a_{0}}{a}\right)^{2}\frac{1}{\left(1+\epsilon\right)}= & \left(1+z\right)^{2}\frac{1+\frac{\epsilon^{2}}{2\left(1+\epsilon\right)}}{1+\left[1+\frac{\epsilon_{0}}{2}\right]\epsilon_{0}},\\
\left(\frac{a_{0}}{a}\right)^{2}\frac{1+\epsilon_{0}}{1+\epsilon}= & \left(1+z\right)^{2}\frac{1+\frac{\epsilon^{2}}{2\left(1+\epsilon\right)}}{1+\frac{\epsilon_{0}^{2}}{2\left(1+\epsilon_{0}\right)}}.
\end{align}
Introducing, with Equation~\eqref{eq:deltaDot}, the~anisotropic perturbation
parameter's derivative, and~applying the expressions above, we can
rewrite with redshift some of the factors present in Equation~\eqref{eq:Hubble(delta,a)=00003D00003DH(z)}
\begin{align}
\frac{\dot{\epsilon}}{3\left(1+\epsilon\right)}= & \frac{\dot{\epsilon}_{0}}{3}\left(1+z\right)^{3}\times\nonumber \\
 & \frac{1+\frac{\epsilon^{2}}{2\left(1+\epsilon\right)}}{1+\left[1+\frac{\epsilon_{0}}{2}\right]\epsilon_{0}}\sqrt{\frac{1+\left[1+\frac{\epsilon}{2}\right]\epsilon}{1+\left[1+\frac{\epsilon_{0}}{2}\right]\epsilon_{0}}},\\
\left(\frac{a}{a_{0}}\right)^{3}= & \left(1+z\right)^{-3}\left(\sqrt{\frac{1+\left[1+\frac{\epsilon}{2}\right]\epsilon}{1+\left[1+\frac{\epsilon_{0}}{2}\right]\epsilon_{0}}}\right)^{-3},\\
\left(\frac{a_{0}}{a}\right)^{3}\frac{1+\epsilon_{0}}{1+\epsilon}= & \left(1+z\right)^{3}\frac{1+\frac{\epsilon^{2}}{2\left(1+\epsilon\right)}}{1+\frac{\epsilon_{0}^{2}}{2\left(1+\epsilon_{0}\right)}}\sqrt{\frac{1+\left[1+\frac{\epsilon}{2}\right]\epsilon}{1+\left[1+\frac{\epsilon_{0}}{2}\right]\epsilon_{0}}}.
\end{align}
They can then be linearized in $\epsilon$ as 
\begin{align}
\frac{\dot{\epsilon}}{3\left(1+\epsilon\right)}\simeq & \frac{\dot{\epsilon}_{0}}{3}\left(1+z\right)^{3}\left(1+\frac{\epsilon_{0}}{2}\left[\frac{\epsilon}{\epsilon_{0}}-3\right]\right),\label{eq:ddelta/1+delta}\\
\left(\frac{a}{a_{0}}\right)^{3}\simeq & \left(1+z\right)^{-3}\left(1-\frac{3}{2}\epsilon_{0}\left[\frac{\epsilon}{\epsilon_{0}}-1\right]\right)\simeq\left(1+z\right)^{-3},\label{eq:aRedshift}\\
\left(\frac{a_{0}}{a}\right)^{3}\frac{1+\epsilon_{0}}{1+\epsilon}\simeq & \left(1+z\right)^{3}\left(1+\frac{\epsilon_{0}}{2}\left[\frac{\epsilon}{\epsilon_{0}}-1\right]\right).\label{eq:denScaling}
\end{align}

\subsection{Anisotropic Hubble Parameter}

The Hubble parameter (\ref{eq:Hubble(delta,a)=00003D00003DH(z)}),
joined to the expressions above, also using Equation~\eqref{eq:dd0/H0},
then reads, to linear order
\begin{eqnarray*}
\mathcal{H}\simeq & \sqrt{\left(\frac{\dot{\epsilon}_{0}}{3}\right)^{2}\left(1+z\right)^{6}\left(1+\epsilon_{0}\left[\frac{\epsilon}{\epsilon_{0}}-3\right]\right)+H_{0}^{2}\left(\Omega_{m}\left[\left(1+z\right)^{3}\left(1+\frac{\epsilon_{0}}{2}\left[\frac{\epsilon}{\epsilon_{0}}-1\right]\right)-1\right]+1+\frac{2}{3H_{0}}\frac{\dot{\epsilon}_{0}}{1+\epsilon_{0}}\right)}\\
\simeq & H_{0}\sqrt{\left(\frac{\epsilon_{0}}{3\Delta_{0}}\right)^{2}\left(1+z\right)^{6}\left(1+\epsilon_{0}\left[\frac{\epsilon}{\epsilon_{0}}-3\right]\right)+\left(\Omega_{m}\left[\left(1+z\right)^{3}\left(1+\frac{\epsilon_{0}}{2}\left[\frac{\epsilon}{\epsilon_{0}}-1\right]\right)-1\right]+1+\frac{2}{3}\frac{\epsilon_{0}}{\Delta_{0}}\right)}
\end{eqnarray*}
and is further reduced to linear order as 
\begin{align}
\mathcal{H}\simeq & H_{0}\sqrt{\Omega_{m}\left[\left(1+z\right)^{3}\left(1+\frac{\epsilon_{0}}{2}\left[\frac{\epsilon}{\epsilon_{0}}-1\right]\right)-1\right]+1+\frac{2}{3}\frac{\epsilon_{0}}{\Delta_{0}}}\label{eq:HubbleLinOrder}
\end{align}
\begin{align}
 & \textrm{with }\Delta_{0}=\frac{2}{\Omega_{m}}\left[\sqrt{\Omega_{m}\left(\left(\frac{a_{i}}{a_{0}}\right)^{-3}-1\right)+1}-1\right].\label{eq:DeltaDef}
\end{align}
From Equation~\eqref{eq:AnisoDepSolNL}, in the general case, we get 
\begin{align*}
\epsilon= & \frac{2\dot{\epsilon}_{0}}{H_{0}\left(\Omega_{m}-x\right)}\left[\sqrt{\left(1-\Omega_{m}+x\right)+\left(\Omega_{m}-x\right)\left(\frac{a_{i}}{a_{0}}\right)^{-3}}\right.\\
 & \left.-\sqrt{\left(1-\Omega_{m}+x\right)+\left(\Omega_{m}-x\right)\left(\frac{a}{a_{0}}\right)^{-3}}\right].
\end{align*}
Then, employing Equation~\eqref{eq:dd0/H0}, as in the derivation from
Appendix \ref{part:Appendix-A:-Solutions}, we linearize in $x$ 
\begin{multline*}
\epsilon\simeq\frac{\epsilon_{0}}{\Delta_{0}}\frac{2}{\Omega_{m}}\left[\sqrt{\Omega_{m}\left(\left(\frac{a_{i}}{a_{0}}\right)^{-3}-1\right)+1}\vphantom{\left\{ \frac{\left(\frac{a}{a_{0}}\right)^{-3}}{2\sqrt{\left(\left(\frac{a}{a_{0}}\right)^{-3}\right)}}\right\} }\right.-\sqrt{\Omega_{m}\left(\left(\frac{a}{a_{0}}\right)^{-3}-1\right)+1}\\
\left.+\left\{ \frac{\left(\left(\frac{a}{a_{0}}\right)^{-3}-1\right)}{2\sqrt{\Omega_{m}\left(\left(\frac{a}{a_{0}}\right)^{-3}-1\right)+1}}-\frac{\left(\left(\frac{a_{i}}{a_{0}}\right)^{-3}-1\right)}{2\sqrt{\Omega_{m}\left(\left(\frac{a_{i}}{a_{0}}\right)^{-3}-1\right)+1}}+\frac{1}{\Omega_{m}}\right\} x+O(x^{2})\right].
\end{multline*}

Finally, introducing redshift with Equation~\eqref{eq:aRedshift} and the
definition (\ref{eq:DeltaDef}), we can obtain the linear form for
$\epsilon$: 
\begin{align}
\epsilon\simeq & \frac{\epsilon_{0}}{\left[\sqrt{\Omega_{m}\left(\left(1+z_{i}\right)^{3}-1\right)+1}-1\right]}\left[\sqrt{\Omega_{m}\left(\left(1+z_{i}\right)^{3}-1\right)+1}-\sqrt{\Omega_{m}\left(\left(1+z\right)^{3}-1\right)+1}\right]\\
\simeq & \epsilon_{0}\frac{\left[1-\sqrt{\frac{\Omega_{m}\left(\left(1+z\right)^{3}-1\right)+1}{\Omega_{m}\left(\left(1+z_{i}\right)^{3}-1\right)+1}}\right]}{\left[1-\frac{1}{\sqrt{\Omega_{m}\left(\left(1+z_{i}\right)^{3}-1\right)+1}}\right]},\label{eq:AnisoDepSolLin}
\end{align}
from which we deduce the ratio 

\begin{align}
\frac{\epsilon}{\epsilon_{0}}\simeq & \frac{\left[1-\sqrt{\frac{\Omega_{m}\left(\left(1+z\right)^{3}-1\right)+1}{\Omega_{m}\left(\left(1+z_{i}\right)^{3}-1\right)+1}}\right]}{\left[1-\frac{1}{\sqrt{\Omega_{m}\left(\left(1+z_{i}\right)^{3}-1\right)+1}}\right]},
\end{align}
We can then use the scale-redshift approximation (\ref{eq:aRedshift})
and the values at recombination to express
\begin{align*}
\left(\frac{a_{i}}{a_{0}}\right)^{3}\simeq & \left(1+z_{i}\right)^{-3}=10^{-3\left(n+3\right)}
\end{align*}
\begin{align*}
\epsilon_{0}\simeq & \frac{\left[\sqrt{\Omega_{m}\left(10^{3\left(n+3\right)}-1\right)+1}-1\right]\epsilon_{r}}{\left[\sqrt{\Omega_{m}\left(10^{3\left(n+3\right)}-1\right)+1}-\sqrt{\Omega_{m}\left(10^{9}-1\right)+1}\right]},\\
\Rightarrow\epsilon\simeq & \frac{\left[\sqrt{\Omega_{m}\left(10^{3\left(n+3\right)}-1\right)+1}-\sqrt{\Omega_{m}\left\{ \left(1+z\right)^{3}-1\right\} +1}\right]\epsilon_{r}}{\left[\sqrt{\Omega_{m}\left(10^{3\left(n+3\right)}-1\right)+1}-\sqrt{\Omega_{m}\left(10^{9}-1\right)+1}\right]},
\end{align*}
(where the relevant input measurements are $\epsilon_{r}\sim10^{-5}$
for $\frac{a_{r}}{a_{0}}=10^{-3}$ and $n=\log_{10}\left(\frac{a_{r}}{a_{i}}\right)$)
to input in Equation~\eqref{eq:HubbleLinOrder} and finally get the Hubble
redshift relation 
\begin{align}
\mathcal{H}\simeq & H_{0}\sqrt{\Omega_{m}\left(\left(1+z\right)^{3}\left\{ 1+\frac{\epsilon_{r}}{2}\frac{1-I\left(z\right)}{I_{0}-I\left(z_{r}\right)}\right\} -1\right)\vphantom{+\left\{ 1+\frac{2\Omega_{m}}{3\sqrt{1-\Omega_{m}}}\frac{\epsilon_{0}}{I_{0}}\right\} }},\nonumber \\
 & \overline{+\left\{ 1+\frac{\Omega_{m}}{3}\frac{\epsilon_{r}}{I_{0}-I\left(z_{r}\right)}\right\} },\label{eq:H(z)}\\
\textrm{with }I_{0}= & \sqrt{\Omega_{m}\left(10^{3\left(n+3\right)}-1\right)+1},\\
I\left(z_{r}\right)= & \sqrt{\Omega_{m}\left(10^{9}-1\right)+1},\\
\textrm{and }I\left(z\right)= & \sqrt{\Omega_{m}\left(\left(1+z\right)^{3}-1\right)+1}.
\end{align}

\reftitle{References}

\end{document}